\documentclass{article}
\usepackage{authblk}
\usepackage{graphicx} 
\usepackage{xcolor}
\usepackage{amsmath}
\usepackage[font=small]{caption}
\usepackage{hyperref}
\usepackage{multirow}
\usepackage{array}
\usepackage{makecell}
\usepackage{hyperref}
\hypersetup{
    colorlinks=true,
    citecolor=red,
    linkcolor=cyan,
    filecolor=red,
    urlcolor=magenta,
}
\usepackage[normalem]{ulem}

\title{Efficient dispersal of submicron solid particles for stratospheric aerosol injection}

\author[1,*]{Yair Segev}
\author[1]{Eitan Y.~Levine}
\author[1]{Yair Bar-Yoseph}
\author[1]{Ori Amsallem}
\author[2]{Yuval Dagan}
\author[1]{Elad Laor}
\author[1]{Shai Rahamim}
\author[1]{Alon Luski}
\author[1]{Eran Daniel}
\author[3]{Eshani Hettiarachchi}
\author[1]{Amyad Spector}
\affil[1]{Stardust Labs, Ness Ziona, Israel}
\affil[2]{Faculty of Aerospace Engineering, Technion -- Israel Institute of Technology, Haifa, 32000, Israel}
\affil[3]{Department of Chemistry and Biochemistry, University of California San Diego, La Jolla, California 92093, United States}
\affil[*]{Corresponding author: \href{mailto:y.segev@stardust-initiative.com}{y.segev@stardust-initiative.com}}

\date{\today}

\usepackage[
    backend=biber,
    style=numeric-comp,
    sorting=none,
    doi=false,url=false,eprint=false]{biblatex}
\renewbibmacro{in:}{}
\newbibmacro{string+doiurl}[1]{%
  \iffieldundef{doi}{%
    \iffieldundef{url}{%
      #1%
    }{%
      \href{\thefield{url}}{#1}%
    }%
  }{%
    \href{\thefield{doi}}{#1}%
  }%
}
\renewbibmacro{volume+number+eid}{%
    \printfield{volume}%
    \setunit{\addcomma\space}%
    \printfield{eid}}
\DeclareFieldFormat{title}{\usebibmacro{string+doiurl}{\mkbibemph{#1}}}
\DeclareFieldFormat[article,inproceedings,techreport]{title}{\usebibmacro{string+doiurl}{#1}}
\DeclareFieldFormat[article]{volume}{\mkbibbold{#1}}

\renewcommand{\cite}[1]{\supercite{#1}}

\begin{document}

\maketitle

\begin{abstract}
    Stratospheric aerosol injection (SAI) using solid particles has been proposed as an alternative to sulfate aerosols for solar radiation modification, but practical deployment faces challenges in efficiently deagglomerating and dispersing powders as submicron particles. Here we experimentally demonstrate pneumatic dispersal of particles in optically optimal size ranges for SAI. Using spherical amorphous silica particles, we find that applying a hydrophobic surface treatment substantially improves dispersibility, with 50-85\% of treated particle mass achieving submicron sizes compared to 10\% for untreated particles. We compare the dispersal of treated particles of different sizes and find that 300 nm particles provide superior deagglomeration than 500 nm particles for the same air consumption. Theoretical modeling of the adhesion forces between particles, combined with surface roughness parameters extracted from atomic force microscopy, successfully predicted the relative dispersibility across different particle types. The pneumatic dispersal system achieved optimal performance at air-to-powder mass ratios of about 10:1. Using the measured dispersed particle sizes, we provide a scaling analysis suggesting that a feasibly sized fleet of dispersal aircraft could provide an aerosol layer sufficient for meaningful climate intervention. These results demonstrate that hydrophobic surface treatment and pneumatic dispersal can overcome the agglomeration challenge for SAI with solid particles.
\end{abstract}

\section{Introduction}
The ongoing rise in global temperatures and the associated extreme weather events motivate the search for climate intervention options to mitigate the worst effects of climate change and buy time for decarbonization and adaptation. Stratospheric aerosol injection (SAI), the concept of dispersing a layer of particles to reflect a portion of the sun’s light\cite{budyko1977climatic,dickinson1996climate,Caldeira2000,crutzen2006albedo}, has attracted increasing interest, as the plausibility of achieving substantial cooling is supported by observations following large volcanic eruptions, which enhance the Junge layer of reflective, submicron-sized sulfate aerosols \cite{mccormick1995atmospheric,Robock}. Many studies have focused on identifying and bounding various risks and uncertainties associated with SAI, including “side-effects” on the ozone layer, precipitation patterns and cloud formation\cite{tilmes2008sensitivity,ferraro2014weakened,wunderlin2024side}. However, global-scale geoengineering also involves practical challenges, such as the prerequisite technology, manufacturing and logistic capacities to disperse an estimated millions of tons per year of submicron particles into the lower stratosphere to achieve a significant global average cooling impact of the order of a few watts per square meter \cite{mcclellan2012cost,smith2018stratospheric}.

Inspired by volcanoes, a large number of studies have discussed the option of deploying sulfate aerosols for SAI. Several recent papers have studied the practicality of sulfate deployment, including the possibility of introducing $\text{SO}_{\text{2}}$ instead of directly deploying an aerosolized sulfate solution or sulfuric acid\cite{bingaman2020stratospheric,smith2022subpolar}. These techniques assume subsequent natural conversion into sulfate droplets, and growth of aerosols via condensation or coagulation to optically effective sizes. However, sulfate aerosols are known to have several unfavorable properties, including significant absorption in the atmospheric longwave window\cite{wunderlin2024side}, heterogeneous reactivity with stratospheric nitrogen oxide reservoirs\cite{hofmann1989ozone,rodriguez1991role}, and impacts on precipitation acidity\cite{visioni2020goes}. These drawbacks have motivated a growing number of proposals for alternative materials for SAI, largely based on minerals, \emph{e.g.} titania, alumina, calcite, silica or diamonds\cite{teller1996global,pope2012stratospheric,weisenstein2015solar,keith2016stratospheric,dykema2016improved,tang2016heterogeneous}. 

Using alternative materials opens the door to selecting or tailoring preferable optical and chemical properties, while possibly harnessing existing manufacturing capabilities for several such materials. Such solid particles would also need to be dispersed with typical sizes of several hundreds of nanometers to efficiently scatter visible light and maintain a significant gravitational lifetime in the lower stratosphere\cite{junge1961stratospheric}. However, dispersal of submicron solid particles is challenging due to the dominance of adhesive forces over inertia at these scales. Based on this understanding and a quantitative analysis, a recent paper \cite{SAIConcernsGernot2025} predicted that practical dispersal systems for proposed mineral aerosols could produce prohibitively inefficient distributions of agglomerates. These larger particles have poor optical scattering efficiencies, necessitating far greater injected quantities than would nominally be needed for optimally-sized particles. For a desired cooling effect, a higher mass requirement could directly result in increased side effects related to increased total aerosol volume (\emph{e.g.}, heating from infrared absorption leading to impacts on atmospheric dynamics and composition), surface area (\emph{e.g.}, heterogeneous reactions with stratospheric trace gases leading to ozone depletion), and particle number concentrations (\emph{e.g.}, enhanced nucleation of cirrus clouds from settling particles), as well as larger costs and burdens on manufacturing capacities, supply chains and the deployment fleet. For substantially non-optimal sizes (on the order of several microns), infrared absorption will surpass shortwave scattering, and the net cooling effect of most particles will become negative\cite{lacis1992climate}. Although several aerosol systems have demonstrated dispersal of submicron mineral particles \cite{werle1974powder}, the few reported attempts to disperse candidate SAI materials in lab settings have generally confirmed the difficulties of achieving optically efficient particle size distributions (PSD)\cite{neukermans2021methods}.

Here we present the efficient dispersal of solid, submicron particles designed for SAI research. Our experimental dispersal method comprises pneumatic eductors, which use pressurized air to convey and deagglomerate fine powders. Several powder samples were tested, with substantial improvements in dispersibility resulting from the application of a hydrophobic surface treatment to the particles. Employing a combination of optical spectrometry and laser diffraction imaging instruments, we measure high mass fractions of particles dispersed in desired size ranges which are near optimal for scattering sunlight. Analysis of the air consumption required for effective dispersal suggests that this pneumatic dispersal method is scalable within the practical constraints of aircraft, and may thus be an enabling building block for SAI technology. While all powders studied here were based on spherical amorphous silica particles, our findings are generally applicable for engineered solid particles with similarly narrow grain size distributions and non-polar surfaces.

\section{Methods}

\subsection{Materials}
All particle samples used in this study were amorphous silica spheres. Amorphous silica has recently been proposed as a building block for SAI particles \cite{Amyad} due to its established safety\cite{safety}, relative chemical inertness, including towards stratospheric gases \cite{Anais2026}, and existing large-scale manufacturing infrastructure \cite{Tamir}. The spheres (``monomers'') were fabricated using a Stöber sol-gel method\cite{stober1968controlled}, a widely tunable mass-production process. For our purposes, the spheres were produced with target diameters of 300 or 500 nm, denoted with the prefix D300 or D500, respectively, as this size range is the most efficient in scattering sunlight. Typical diameter spreads of 10-15\% (one standard deviation) were measured using scanning electron microscopy (SEM). Based on SEM images and dynamic light scattering (DLS) measurements, the mass fraction of ultra-fine particles, \emph{i.e.} with diameters smaller than 100 nm, was less than 0.1\%.

Untreated particles, denoted with the suffix U, were strongly hydrophilic due to dense coverage by hydroxyl (OH) groups typical of silica surfaces, as confirmed by atomic force microscopy-photothermal infrared (AFM-PTIR) spectroscopy (see section \ref{surface characterization} in the Supplementary Materials). Surface-treated samples were subjected to silanization\cite{SilanizationArkles1977}, whereby trimethylsilyl (TMS) functional groups were attached to replace or sterically hinder hydroxyl groups. Following treatment, these particles were strongly hydrophobic, with typical contact angles of 130$^\circ$ with water. 

For increased flowability, some hydrophobic samples, denoted with the suffix M, were mixed with additional nanometric silica spheres in a mechanofusion process. These fused spheres, with typical diameters of around 50 nm, constitute only about 0.1\% of the powder mass, and act as spacers between the larger monomers. 

To avoid the effects of hygroscopicity on dispersal, which was especially pronounced for untreated hydrophilic samples, all powders were dried in an oven at 155$^\circ$C for 15 hours before each experiment session.

Table 1 summarizes the four samples discussed in this study, including key values of the mass (or volume) distributions of their monomers. More information about these particles and their production is provided elsewhere\cite{Amyad, Tzemah}.

\begin{table}[!b]
\hbadness=10000
\small
\resizebox{\textwidth}{!}{%
\begin{tabular}{|l|p{6.2cm}|c|c|c|}
\hline
\textbf{\multirow{2}{*}{Name}} & \textbf{\multirow{2}{*}{Description}} & \multicolumn{3}{c|}{\textbf{Monomer mass distribution}} \\
\cline{3-5}
 &  & \text{D$_{10\%}$} & \text{D$_{50\%}$} & \text{D$_{90\%}$} \\
\hline
D300 & Surface-treated 300 nm spheres & 275 nm & 310 nm & 355 nm \\
\hline
D500 & Surface-treated 500 nm spheres & \multirow{3}{*}{440 nm} & \multirow{3}{*}{500 nm} & \multirow{3}{*}{565 nm} \\
\cline{1-2}
D500U & Untreated 500 nm spheres &  &  &  \\
\cline{1-2}
D500M & Surface-treated 500 nm spheres with mechanofused nanoparticles (0.1\% by mass) &  &  &  \\
\hline
\end{tabular}%
}
\caption{Materials used in this study. All materials were amorphous silica produced by a Stöber sol-gel method. Surface treated samples were silanized to reduce adhesion. The values \text{D$_{10\%}$}, \text{D$_{50\%}$} and \text{D$_{90\%}$} are the diameters corresponding to cumulative fractions of 10\%, 50\% and 90\% of the monomer mass distributions, respectively.}
\label{tab:materials}
\end{table}

\subsection{Experimental setup}
\begin{figure}[!tb]
    \centering
    \includegraphics[width=1\linewidth]{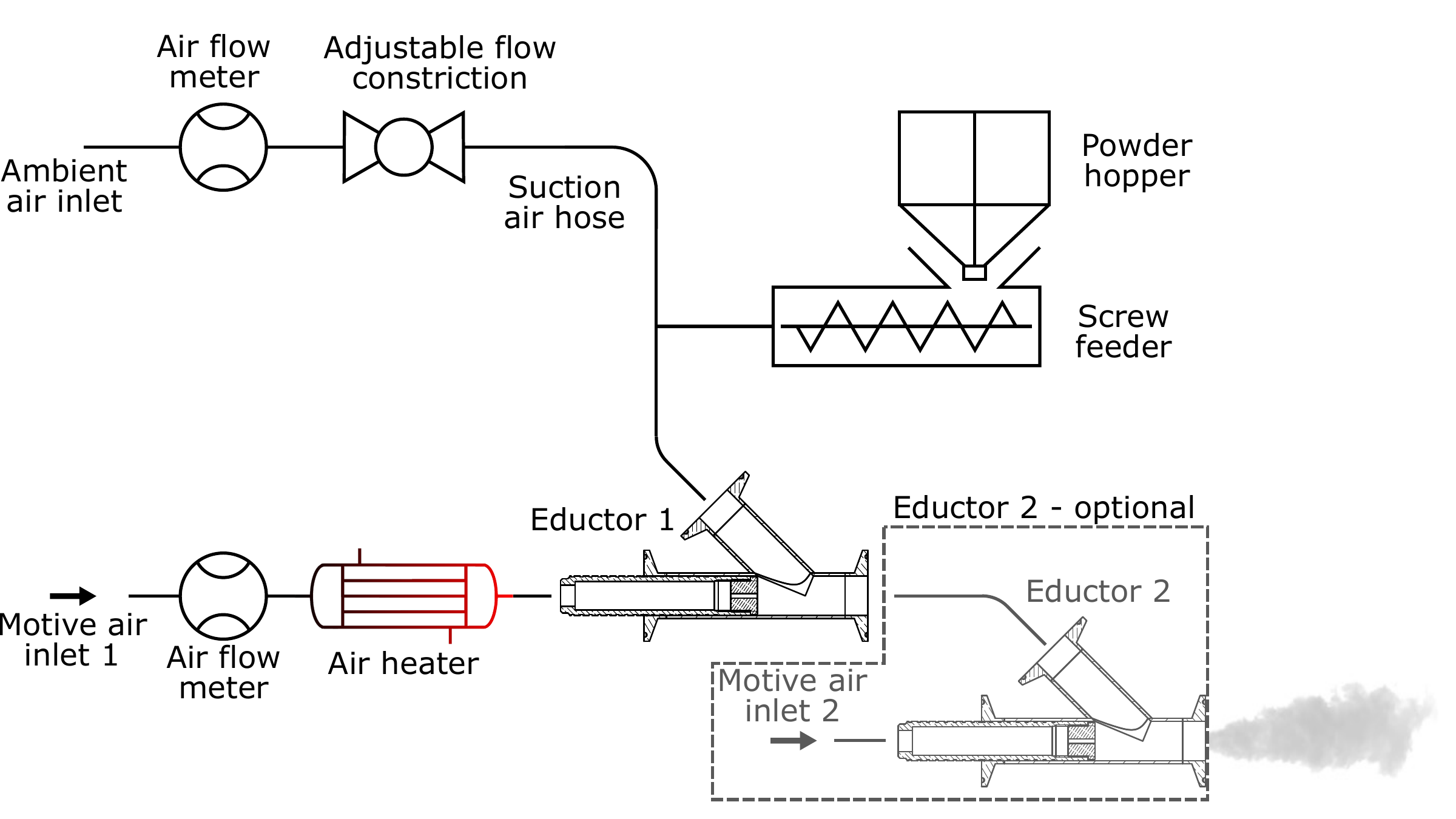}
    \caption{Schematics of the dispersion setup. Pressurized motive air enters the eductor through a controlled heating element, and expands through an orifice. Powder is dosed from a hopper, and sucked into the eductor by the expanding motive air. An adjustable constriction on the suction hose enables optimization of the air consumption. An optional second eductor stage can be used to improve deagglomeration. Detailed schematics of the eductor are provided in Fig.~\ref{fig:eductor} in the Supplementary Materials. }
    \label{fig:system}
\end{figure}

The basic form of the experimental dispersal setup, described in Fig.~\ref{fig:system}, includes a compressed (“motive”) air source, a Venturi eductor with a choked-flow nozzle through which the motive air expands, a suction inlet through which the powder is pulled in, a mixing volume where the two flows mix and turbulent shear stresses evolve, and an exhaust which discharges the plume of mixed air and powder. The material samples are fed from a hopper into the suction hose using a dosing mechanism (Coperion K-Tron K-MV-KT20 volumetric twin-screw feeder). The motive and suction air flow rates are monitored using thermal mass flow meters (Festo SFAB-600U and Ningbo KF100, respectively). The cross-section of the ambient-air side inlet of the suction hose can be constricted to minimize the total air consumption while maintaining sufficient flow to transport the powder. The motive air can be heated before expansion with an inline heater (Farnam Cool Touch 150). All pressure values stated in this paper refer to absolute pressure.

For improved deagglomeration, an optional second eductor was connected in sequence, with the exhaust from the first feeding into the suction inlet of the second. This configuration provided pre-aeration of the powder before the final turbulent mixing stage. The main dispersal system parameters investigated in this study included air pressure and temperature to the eductors, and particle mass feed rates. Detailed schematics of the eductors used in this study are provided in Fig.~\ref{fig:eductor} in the Supplementary Materials. 

The dispersal system was placed at one end of a dedicated “spray tunnel” facility, depicted in Fig.~\ref{fig:spray tunnel} in the Supplementary Materials, with remote control of all systems from an adjacent control room. Optical particle sampling instruments, detailed below, were placed at the far end, up to 4 meters downstream of the dispersal system. The tunnel infrastructure included circulation through multiple filter stages at a rate of up to 3 exchanges per minute, to minimize buildup of particle-laden air near the instruments and to allow a prompt return to clean, monitored background aerosol levels after each experiment. The tunnel was designed to be wide enough to avoid interactions between the plume and the walls or floor before the particles arrive at the sampling area, and the plume flight time was short enough to avoid gravitational stratification of the PSD for particles in the entire range of interest. Typical experiments lasted 5-10 seconds, much longer than the stabilization times of the flows and pressures in the dispersal system. Maximum feed rates that could be sustained in this facility are on the order of tens of kilograms per hour.

The sampling setup included two complementary instruments: an optical particle spectrometer (OPS; GRIMM 11-D), and a laser diffraction imaging instrument (LDI; Malvern Spraytec, 300 mm lens, 1400 mm bench). The OPS provided high resolution particle spectra for particles with optical diameters of 0.3 to 10 microns, at up to 1 Hz. To avoid saturation effects, the OPS sampled air through a calibrated, 1:100  dilution system (Topas DIL 557). 

The LDI was employed to extract the partitioning of the dispersed load between “small” (up to 10 $\mathrm{\mu m}$) and large (10 to 1000 $\mathrm{\mu m}$) agglomerates, as the high, kHz-scale detection rate of this instrument allows sampling of low-incidence high-mass particles. The instrument was placed in a custom-built enclosure with active suction to avoid contamination, with air curtains protecting the optics and laser source windows. A custom skimming aperture was attached in front of the beamline to select a portion of the plume, allowing control of the optical depth to maintain sufficient signal-to-noise ratio of scattered light while avoiding multi-scattering effects at lower transmission. 

The sampled spectra of the two instruments were merged to benefit from their complementary advantages, and the particle size bins were corrected to account for the measured refractive index of the silica particles. All particle diameters presented in the measured spectra refer to optical diameters, which is the diameter of a spherical particle with the same refractive index that would register with the measured signal in our setup. For monomers, this optical diameter is equivalent to the geometric diameter of the sphere, and for a compact (roughly spherical) agglomerate of monomers this diameter will approximately correspond to that of a sphere with equal volume. The scattering cross-section of such compact agglomerates for sunlight will also roughly correspond to spheres of the same optical diameter defined here \cite{Yoav}.

SEM images of agglomerates emitted from the dispersal system, and collected with a cascade impactor (TSI mini-MOUDI 135), showed that relatively compact shapes were prominent for larger agglomerates. Rarer, strongly elongated aggregates that are much larger than the laser wavelengths (\emph{i.e.}, far enough from resonance effects) will likely register with some statistically sampled spread around their volumetric equivalent diameter, due to their random orientation during sampling. All images of thousands of collected particles showed no signs of any particle fracturing; therefore, the smallest dispersed particles are monomers. 

As all particles in this study have the same true density, the results will be presented in mass fractions instead of volume. Particle size distributions (PSDs) refer to the normalized logarithmic mass density, 
\begin{equation} \label{eq:PSD}
\rho(D)=\frac{1}{M}\frac{dm}{d\,\text{log}D}, 
\end{equation}
where $D$ is the particle diameter in microns, $M$ is the total mass dispersed in the measured range, between $D_{min}=0.3\,\mathrm{\mu m}$ and $D_{max}=1000\,\mathrm{\mu m}$, and $m(D)$ is the cumulative mass dispersed and measured between $D_{min}$ and and $D$. We will refer to the normalized cumulative mass distribution function, $m(D)/M$, as CDF.

\subsection{Theoretical model for adhesive forces}
The adhesive force between monomers was calculated using a rigid sphere model modified to include surface roughness and chemical bonds. The dispersive (van der Waals) force between spheres was calculated using the Rabinovich model\cite{rabinovich2000adhesion1}, which includes an experimentally validated treatment of the surface roughness\cite{rabinovich2000adhesion2}; for a smooth sphere of radius $R$ adhering to a planar surface with a roughness characterized by the root mean square amplitude $a_\mathrm{rms}$ and wavelength $\lambda$, this force is:
\begin{equation} 
F_{Rabinovich}=\frac{HR}{6\delta ^2}\left[\frac{1}{1+\frac{58a_\mathrm{rms}R}{\lambda ^2}}+\frac{1}{1+\frac{1.82a_\mathrm{rms}}{\delta}}\right],
\end{equation}
where $H$ is the effective Hamaker constant between the two bodies and $\delta$ is the minimum contact distance related to the Born repulsion cutoff. For two spheres in contact, we modify the equation by replacing $R$ with the effective radius $R^*=R_1R_2/(R_1+R_2)$; thus, for identical spheres, this reduces the force by two.

The Rabinovich model for the dispersive force does not include the effects of chemical bonds, which are substantial for untreated silica\cite{zhuravlev2000surface}, especially for rough surfaces where $a_\mathrm{rms}\gg\delta$. In our case, the Tabor parameter is much smaller than unity, justifying the use of the Derjaguin-Muller-Toporov (DMT) contact model for rigid spheres\cite{derjaguin1975effect,marshall2014adhesive}; avoiding the redundant dispersive portion of the model, we add only the contribution of the surface chemical bonds,
\begin{equation} 
F_D=2\pi R^*\gamma,
\end{equation}
where $\gamma$ is the surface energy density due only to H-bonds from the hydroxyl groups on the surfaces.

Table \ref{tab:model parameters} summarizes the parameters for three materials used in this study -- 300 and 500 nm treated particles, and 500 nm untreated particles. The mechanofused powder is omitted due to large uncertainties in the dimensions of the nanometric spacers.

The bulk Hamaker constant $H$ for amorphous silica, $6.5\times10^{-20}\,\mathrm{J}$, was taken from literature, along with the cutoff distance for untreated silica, $\delta = 0.165\, \mathrm{nm}$, as the values of $H$ are typically extracted with an assumed value of $\delta$\cite{BERGSTROM1997125,marshall2014adhesive}. The surface roughness parameters for both 500 nm particles were extracted using height images collected by atomic force microscopy (AFM). We assumed that $a_\mathrm{rms}$ for surface treated particles remains the same for the smaller monomers, and that $\lambda$ scales with the diameter of the particles. For treated surfaces, $\delta$ was taken as $0.3\,\mathrm{nm}$, to account for the typical steric hindrance of the TMS groups. For untreated particles, we estimate the probable range of the surface energy density $\gamma$ by accounting for the OH density of moderate to highly hydroxylated amorphous silica surfaces \cite{zhuravlev2000surface} multiplied by the H-bond energy of silanol groups\cite{Musso2017}. For treated particles, we assumed no chemical contribution; this choice is justified by observing a lack of hydroxyl groups on these surfaces (using AFM-PTIR chemical maps), and the fact that TMS groups have extremely low polarizability. Additional information on the acquisition of each parameter is provided in Supplementary Material section \ref{surface characterization}.

Table \ref{tab:model parameters}, also presents the calculated adhesion force for these three materials. The adhesion between 500 nm treated particles is approximately 3 times stronger than for 300 nm particles. For the untreated particles, adhesion is stronger by an additional factor of about 20. As will be shown, for identical dispersal conditions, the efficiency of deagglomeration will correlate accordingly.

\begin{table}
    \centering
\hbadness=10000
\small
\begin{tabular}{|l|c|c|c|}
\hline
\textbf{Sample} & \textbf{D300} & \textbf{D500} & \textbf{D500U} \\\hline Roughness,   $a_\mathrm{rms}$& [3.0, 4.5] nm& [3.0, 4.5] nm& [1.9, 2.7] nm\\ \hline
Roughness   wavelength, $\lambda$ & [18, 36] nm& [30, 60] nm& [30, 60] nm\\ \hline
Contact   distance, $\delta$ & 0.3 nm & 0.3  nm & 0.165 nm \\\hline H-bond   surface energy, $\gamma$& 0 & 0 & [10, 50] \text{mJ/m$^2$}\\\hline \textbf{Total   adhesion force}& \textbf{[0.16, 0.84] nN}& \textbf{[0.42, 2.2] nN}& \textbf{[10, 49] nN}\\ \hline
\end{tabular}
\caption{Particle parameters used for the model, and their associated maximum uncertainty ranges.}
\label{tab:model parameters}
\end{table}

\subsection{Error analysis}
Particle size distributions were assessed by compiling particle number spectra during the entire duration of six repeated experiments per data point. PSD errors account for Poissonian shot noise errors based on total counted particles in each size bin. Systematic errors included corrections for the refractive index of the particles ($n=1.445\pm0.005$ for red light, relevant to both optical instruments, as measured by the Becke line and dispersion staining techniques). Air to powder ratio errors are $\pm5\%$, mainly due to powder feed rate consistency. Particle true density, used when translating particle volume distributions to SAI mass requirements, is $\rho_p=$2.1$\pm$0.1 g/cc, as measured by liquid pycnometry.

\section{Results}
\begin{figure}[!tb]
    \centering
    \includegraphics[width=1.0\linewidth]{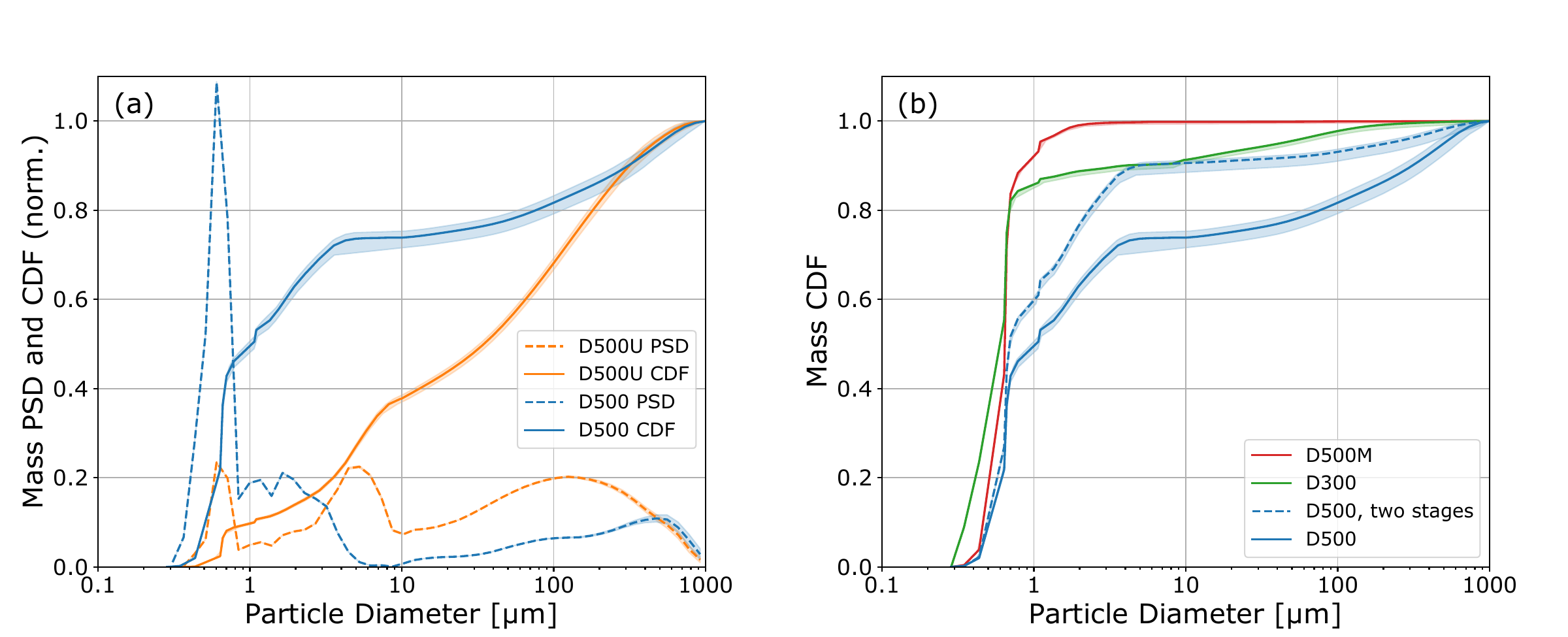}
    \caption{Dispersal of different monomers. (a) Particle size distribution (PSD), representing normalized logarithmic mass density, and cumulative mass distribution function (CDF), for surface-treated (D500) and untreated (D500U) powders of 500 nm diameter silica spheres. Untreated particles present a broad mass distribution ranging three orders of magnitude, with a negligible fraction in the target, submicron range. In contrast, hydrophobic treated particles were predominantly dispersed in size ranges corresponding to monomers or small aggregates. Both measurements were conducted with motive air at 7 bar and 2 kg/hr particle feed rate. (b) Comparison of CDF for various treated samples with the same dispersal parameters as panel (a). Samples with 300 nm monomers (D300) exhibit superior dispersibility than 500 nm particles particles. Mechanofusion of D500 particles with nanometric spheres (D500M) further improves the dispersal, with almost all mass dispersed in the submicron range. Some improvement of D500 dispersal can also be achieved using a two-stage dispersal setup (dashed curve). Here, a low-flux eductor, with a 2 mm nozzle fed at 3 bar, provides aeration before feeding the powder into the primary stage. The second eductor operated with 7 bar and a reduced, 2.2 mm nozzle, to maintain the total motive air consumption of the single-stage measurements. The shaded areas in all plots mark a $\pm1\sigma$ equivalent region.}
    \label{fig:materials}
\end{figure}

The most significant factor enabling efficient dispersal of the silica spheres was the hydrophobic surface treatment. Fig.~\ref{fig:materials}(a) presents examples of particle size distributions for treated (D500) and untreated (D500U) powders with 500 nm monomers using the same dispersal conditions. The treated particle PSD shows a prominent peak at around 500 nm, with a mostly decaying tail of agglomerates up to a few microns. Approximately 50\% of the cumulative mass of treated particles is below 1 $\mathrm{\mu m}$, and about $75\%$ is below 3 $\mathrm{\mu m}$. In contrast, only around $10\%$ of the mass of untreated particles is dispersed in the submicron range, with most agglomerates sized at tens of microns. 

The solid curves in Fig.~\ref{fig:materials}(b) demonstrate the effect of different monomer types on dispersibility, for the same dispersal system parameters. All three materials compared here are surface-treated. The solid blue curve is the same CDF from Fig.~\ref{fig:materials}(a), for 500 nm monomers. As can be seen in the green curve, we found that, for these conditions, 300 nm monomers are more readily dispersed in the submicron range than the 500 nm treated particles (85\% of the powder mass, compared to about 50\%). This result may be counterintuitive given that smaller particles are generally less dispersible than large ones. However, for the particles used here, which are almost perfect spheres with nanometer-scale surface roughness, the adhesive force is lower for the smaller monomers; thus, for similar flow stresses, we can expect that their aggregates will be more prone to fragmentation.

\begin{figure}[!tb]
    \centering
    \includegraphics[width=0.75\linewidth]{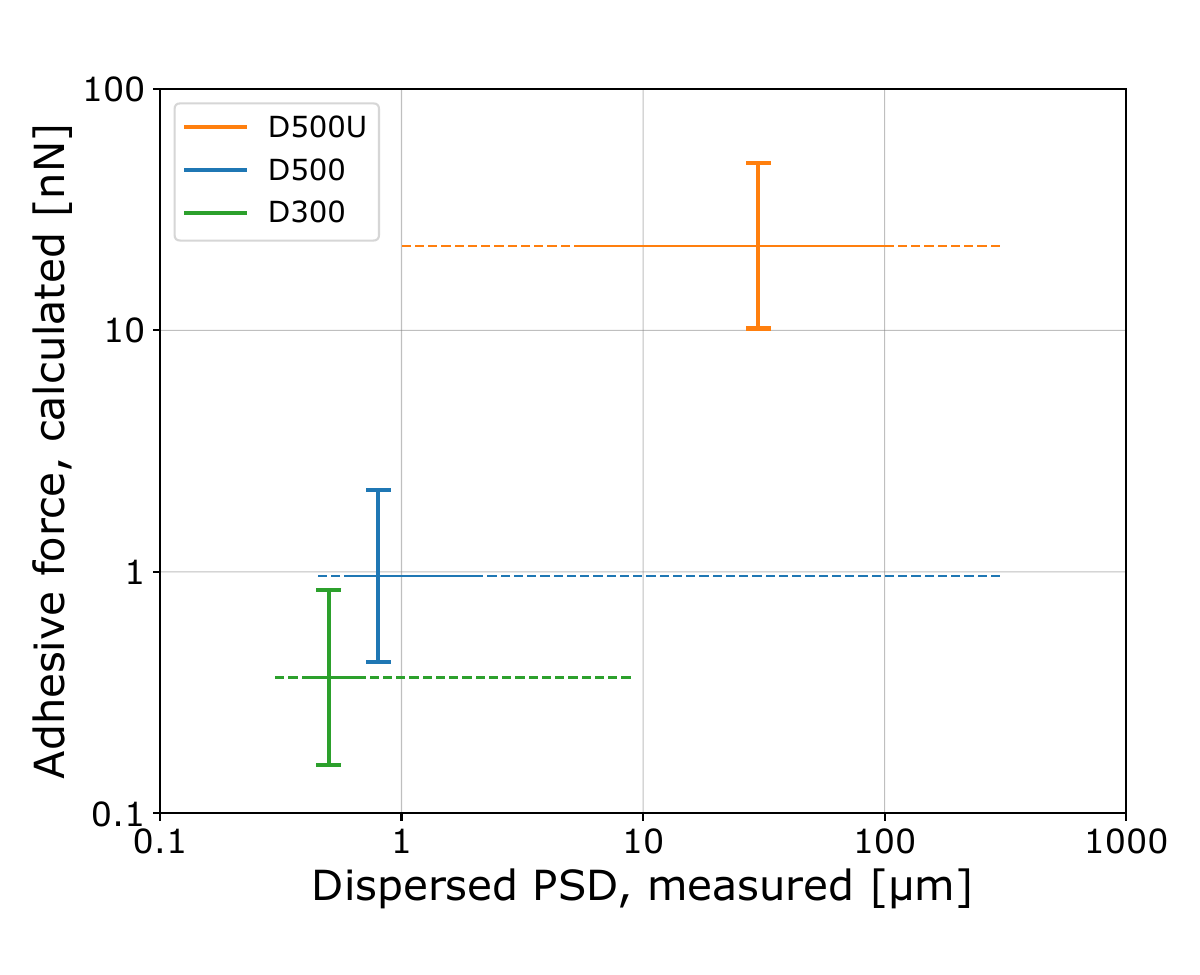}
    \caption{Calculated adhesive force vs. measured, dispersed size distributions for powders with different monomer samples. The vertical ranges correspond to the uncertainty in the model parameters for each material. The horizontal ranges correspond to the CDF; the dashed sections represent the $10^\mathrm{th}-90^\mathrm{th}$ percentile ranges, the solid sections represent the $30^\mathrm{th}-70^\mathrm{th}$ percentile ranges, and the vertical bars are centered on the $50^\mathrm{th}$ percentile. The estimated forces correlate well with the primary size ranges of dispersed particles. Measurements were conducted with motive air at 7 bar and 2 kg/hr particle feed rate. Model parameters and uncertainties are detailed in Table  \ref{tab:model parameters}.}
    \label{fig:adhesion}
\end{figure}

Figure \ref{fig:adhesion} summarizes the measured size distributions and the calculated adhesive force for these materials. The vertical ranges correspond to the uncertainty in the model parameters for each material, from Table \ref{tab:model parameters}. The horizontal ranges correspond to the CDF for each case in  Fig.~\ref{fig:materials}(b); the dashed sections represent the $10^\mathrm{th}-90^\mathrm{th}$ percentile ranges, the solid sections represent the $30^\mathrm{th}-70^\mathrm{th}$ percentile ranges, and the vertical bars are centered on the $50^\mathrm{th}$ percentile. We observe a clear correlation between the estimated forces and the primary size ranges of dispersed particles, indicating that the Rabinovich model, with surface roughness parameters derived from AFM, is a good predictor for relative dispersibility.

Even better deagglomeration was achieved with the samples of 500 nm monomers with mechanofused nanometric spheres (red curve in  Fig.~\ref{fig:materials}(b)), where practically all the mass was dispersed as particles smaller than 3 $\mathrm{\mu m}$, and around $95\%$ below 1 $\mathrm{\mu m}$. This superior dispersibility of mechanofused particles is expected, as the surface asperities caused by the fused nanometric spheres create gaps and lower the contact area between the larger monomers, substantially decreasing the adhesion energy as well as the lubrication forces which resist their separation. We note, however, that the mechanofused nanometric spheres would be considered ultrafine particles if they became dislodged from larger monomers, and might be subjected to additional scrutiny from a respiratory safety perspective. 

Another option for improving the dispersibility of the 500 nm particles, shown in the dashed curve, is to feed the emitted particles directly into a second eductor stage. Here, the first stage eductor is fed with lower pressure air, providing aeration of the powder before the primary deagglomeration stage.

\begin{figure}[!tb]
    \centering
    \includegraphics[width=1.0\linewidth]{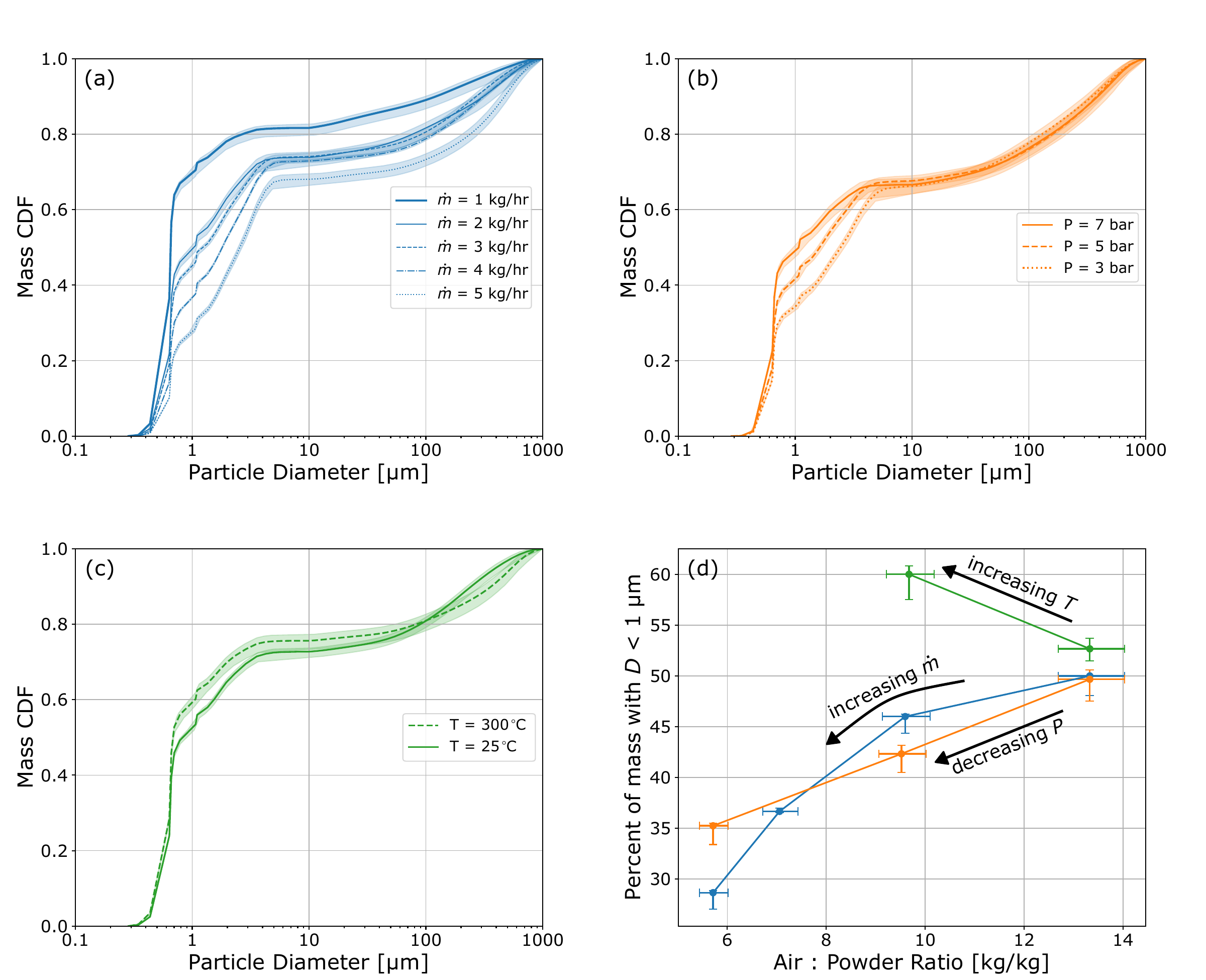}
    \caption{Effect of particle feed rate and air supply variations on dispersal. (a) The cumulative size distributions for five different feed rates with the same dispersal system configuration show a degradation in submicron dispersal efficacy with increasing feed rate, likely attributed to increased post-dispersal coagulation or reduced turbulent shear stresses at high particle to air volume ratios. (b) Reducing the air consumption by decreasing the motive air pressure from 7 bar to 5 bar or 3 bar leads to degraded dispersal efficiency in the submicron range, consistent with the effect of increasing the particle feed rate. (c) In contrast, reduction of the air consumption can be achieved by heating the motive air. At $300^\circ\mathrm{C}$, slight improvement in dispersibility is observed, attributed to the higher air velocity leading to increased shear stresses. (d) The total fraction of mass dispersed in the submicron range, plotted against the mass ratio of motive air to powder feed rate, demonstrates a 20\% gain with 30\% reduction in air  consumption when heating to $300^\circ\mathrm{C}$, in contrast to comparable reductions in the submicron mass fraction when feed rate is increased or pressure is reduced. All samples included treated, 500 nm monomers (D500). Except where stated, motive air was provided at 7 bar.}
    \label{fig:flow-control}
\end{figure}

For most practical applications, including SAI dispersal systems, it is advantageous to maximize the rate of powder dispersal. However, as shown in Fig.~\ref{fig:flow-control}(a), there is a very clear degradation in the dispersed PSD when the powder feed rate is high. For the dispersal system parameters shown in the figure, a five-fold increase in feed rate leads to a threefold reduction in the submicron fraction of the dispersed mass.  

There are several likely causes for this degradation at high feed rates. First, an increased mass concentration in the dispersed plume increases the rate of collisions between particles. If the rate of dilution due to dissipation is not fast enough, and if the sticking probability between colliding particles is high, coagulation (reagglomeration) of the powder may occur, shifting the PSD to higher diameters\cite{Fuchs}. Alternatively, the high mass flux may impact deagglomeration by changing the behavior of the flow in the dispersal system. In particular, when a high fraction of a flow's volume consists of particles, turbulent shear can be strongly subdued. For the air and powder fluxes used here, the air-particle volume ratio is 0.01\% to 0.1\% on average, and likely higher in certain regions, well within regimes that suggest that both effects are significant\cite{turbulence}. To achieve a higher mass flux without PSD degradation, a larger eductor nozzle or higher motive air pressure could be used to increase the flux of air. Alternatively, dispersal rates could be scaled up by running multiple dispersal systems in parallel at low flux each, also requiring more volumes of air.

The use of more air to achieve effective deagglomeration in either solution could limit the achievable dispersal rate, as air supply would eventually be constrained by the energy available from the dispersal platform (see discussion in Section 4). Fig.~\ref{fig:flow-control}(b) presents the effect on PSD of reducing air supply by decreasing the motive air pressure from 7 bar to 5 bar or 3 bar. Some degradation in the PSD is observed, in particular for particles in the range of 0.6 to 1 $\mathrm{\mu m}$, which are mostly shifted to $>1\mathrm{\mu m}$ at 3 bar. As this size range includes the most efficient particles for scattering sunlight, this shift would be highly detrimental for SAI applications.

In contrast, an opposite trend is found when air temperature is varied to conserve air,~\ref{fig:flow-control}(c). Here, the motive air was heated before passing through the nozzle, and the PSD shows a clear improvement with higher temperature. For pressures above twice the exhaust pressure (ambient in the lab, and much lower in the stratosphere), motive air consumption is well described by the choked nozzle equation,
\begin{equation}
\dot{m_a} \;\approx\; 0.538 \; \frac{p_{0} \,d_n^{2}}{\sqrt{R \, T_{0}}},
\end{equation}
where $d_n$ is the nozzle diameter, $R\approx287\mathrm{\frac{J
}{kg \, K}}$ is the specific gas constant for air, and $p_0$ and $T_0$ are the upstream (motive) air pressure and temperature, respectively. The inverse relation of the air consumption with the square root of the temperature is the combination of two counter-acting effects: a decrease in air density with higher temperature and constant pressure, and an increase in the choked (sonic) flow velocity in the nozzle, which is proportional to $\sqrt{T_0}$. We attribute the improvement in deagglomeration to the latter effect -- namely, the higher motive air flow velocity induces stronger shear when mixing with the suction flow carrying the particles, consequently increasing the turbulent stresses on agglomerates. We compared the spectra measure with heated and unheated air in many experimental configurations and parameters, and found improved dispersal in all cases.

Fig.~\ref{fig:flow-control}(d) summarizes the effects of varying feed rate, pressure, and temperature on the mass fraction dispersed in the submicron range. The three cases from panels (a-c) are plotted against the mass ratio of motive air to powder, $a_R$; this parameter will be revisited in the next section as it pertains to the applicability of a dispersal system for aerial platforms. As can be seen, an approximate 30\% reduction in air consumption can be achieved by heating, with a 20\% gain in the submicron mass fraction, in contrast to a degradation of similar magnitude when pressure is reduced or feed rate is increased. We note that hot air may be readily available in aviation systems, including the bleed air systems from the engines. Thus, heated motive air may be a viable option for reducing air consumption in SAI dispersal systems without degrading the PSD.

The dispersal setup in our experiment could be optionally grounded to avoid tribo-electric charging during high powder flows. Using a micro-ampere current sensor attached to the grounding line, we found that D500 particles carry on average 6-7 negative elementary charges per monomer. Over the duration of a typical measurement, no effect on PSD was observed when the grounding line was disconnected, However, an SAI platform dispensing large quantities of powder would likely need to include means for discharging the static charges accumulated from the dispersal system. 

Our experimental setup has demonstrated successful deagglomeration and dispersal of submicron solid particles. Previous estimates by Hack et al.\cite{SAIConcernsGernot2025} for dispersal of SAI powders assumed that particles would be fed through a choked nozzle. Based on another theoretical study \cite{forney1983scaling}, they concluded that higher air-solid mass ratios (35:1) and higher pressures (more than 10 bar) would be required for deagglomeration, compared to our findings. The break-up mechanism in their configuration is stresses on the agglomerates from shock waves in the nozzle's diverging section, whereas in our subsonic configuration the stresses arise from turbulent shear \cite{yao2021deagglomeration}. The more crucial difference, however, stems from the assumption of smooth spheres in their model for the dimensionless Weber number; our implementation of the Rabinovich model shows a substantial reduction in adhesion when the particle surface roughness is taken into account, likely enabling other deagglomeration configurations.

\section{Implications for SAI}

The primary indicator for the efficacy of the dispersed PSD is the amount of cooling achievable per unit mass of powder. The combined effect from both scattered and absorbed light, and possibly higher order effects, is termed the radiative forcing (RF), and, for the sake of the analysis here, RF will always be globally and annually averaged. Since SAI particles will eventually settle out of the stratosphere and need to be replaced, the efficacy will relate the RF to a rate of injection.

\begin{figure}[!tb]
    \centering
    \includegraphics[width=0.7\linewidth]{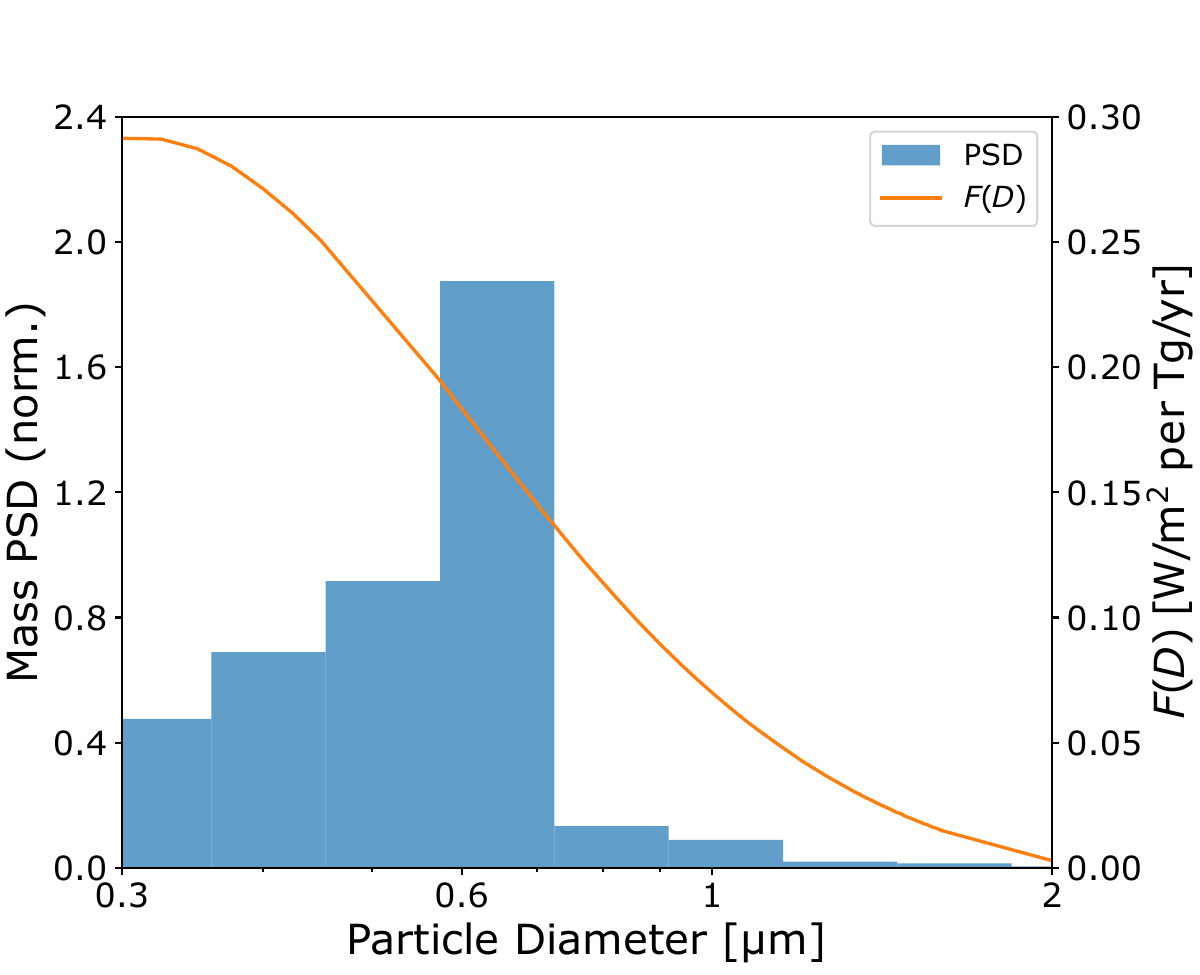}
    \caption{RF mass efficiency $F(D)$ for silica particles for representative deployment parameters\cite{Yoav}, superimposed on a measured particle size distribution of D300 particles. Integrating the overlap of the functions gives $\eta_M=0.2\pm 0.01\,\mathrm{W/m^2}$ per Tg/yr. The PSD, which includes 98\% of the dispersed mass, was measured with a two-stage setup, $\dot{m}=4$ kg/hr, 300 $^\circ$C, 3.6 bar, and $a_R=7.5$ kg/kg.}
    \label{fig:SAI_eff_calc}
\end{figure}

We define the RF mass efficiency, $\eta_M$,  as the globally averaged net reflected irradiance ($\mathrm{W/m^2}$) per annual injection rate,
\begin{equation}
\eta_M=\int F(D)\rho(D)d(\text{log}D),
\end{equation}
where $F(D)$ is the RF contribution of particles with diameter $D$ and $\rho(D)$ defined in Eq.~(\ref{eq:PSD}). Injection rate is typically given in Tg/yr, where Tg signifies teragrams (million tons). The values of $F(D)$ will depend on the details of the dispersal strategy, as well as the total dispersed burden (due to coagulation effects on particle lifetime). A representative function $F(D)$, adopted from Lederer at al.\cite{Yoav}, is plotted in Fig.~\ref{fig:SAI_eff_calc} in orange. For the scenario detailed in Lederer et al., the optimal efficiency of our silica spheres is $F\approx 0.29\,\mathrm{W/m^2}$ per Tg/yr, corresponding to $D\approx300\,\mathrm{nm}$. The calculation includes coagulation, which favors the dispersal of particles which are smaller than optimal for scattering sunlight. For a relatively effective dispersal of D300 particles presented in Fig.~\ref{fig:SAI_eff_calc}, the resulting RF mass efficiency is $\eta_M=0.2\pm 0.01\,\mathrm{W/m^2}$ per Tg/yr; as an illustrative example, this would mean that 10 Tg/yr would be needed to balance the 2020-2050 increase in total anthropogenic effective radiative forcing under the IPCC AR6 high-GHG scenario SSP3-7.0\cite{Dentener2021}. Note that the uncertainty presented for $\eta_M$ only includes contributions from the PSD, and does not include uncertainties in particle transport models or climatic effects.

The effective dispersal of particles as aerosols of several hundred nanometers optical diameter is a prerequisite for SAI. However, to achieve a substantial RF, a practical dispersal system for SAI must enable the deployment of several Tg of submicron particles per year within the constraints of a feasibly sized fleet of aircraft. Here, we provide a rough assessment of these constraints and the compatibility of our system.

First, we note that there are currently no mass-produced aircraft types with the capability of lofting substantial payloads into the lower stratosphere at all latitudes – \emph{i.e.}, at least 60 thousand feet. There have been proposals for conducting SAI at lower altitudes accessible to existing aircraft, \emph{e.g.}, converted business jets confined to dispersal in the lower stratosphere outside the tropics\cite{LowAltitudeSAI}, though it is unclear if this scenario would suffice for global-scale climate intervention. For larger aircraft, converted or developed for SAI missions in the lower stratosphere, we assume characteristics similar to existing cargo aircraft or converted freighters, particularly in terms of net payload capacities of at least 50 tons per flight. To deploy such amounts, a dispersal system such as we have implemented here would have to be scaled up if possible, or multiple systems could be installed  in parallel on the platform.

The dispersal platform would also have to supply air to disperse the powder. While small amounts of air could possibly be provided by standard bleed air from the aircraft engines or existing onboard auxiliary power units (APU), high amounts would likely require the installation of a dedicated turbine engine to supply the motive air, either directly (\emph{i.e.}, from its exhaust) to the dispersal system or by driving another air compressor. As a rough estimate of the air that could be supplied, we note that large aviation engines typically pass about 100~$\mathrm{kg/s}$ of air at high altitude. We found that efficient deagglomeration required an air-to-powder mass ratio $a_R\sim10$. Thus, the platform could dispense its payload over a 1-2 hour leg at dispersal (cruising) altitude, comfortably allowing an order of 500 flights per year. Therefore, for every 1 Tg/yr of particles, approximately 50 aircraft would need to be employed. For $\eta_M=0.2$ $\mathrm{W/m^2}$ per Tg/yr, a reasonable fleet of hundreds of aircraft could provide the aerosol layer needed to reflect up to a few $\mathrm{W/m^2}$.

Although fleet size is a crucial factor, we note that other parameters may be equally or more important considerations when optimizing the dispersal parameters or PSD for a target radiative cooling. For example, optimizing $\eta_M$ would achieve a minimal amount of injected mass for a desired cooling, minimizing manufacturing costs, supply chain burdens, and total payload requirements, while targeting a minimal total surface area of aerosols injected into the stratosphere would help minimize the risks of chemical side-effects. As an example, we consider a specific, measured case where varying the air consumption for a given dispersal system configuration changes the PSD, and thus the radiative forcing per particle mass. In addition to $\eta_M$, we define the following efficiency parameters, which are related in different ways to efficacy, safety and practicality. 

First, we can define the RF surface area efficiency, $\eta_S$, which can be measured in $\mathrm{W/m^2}$ per $\mathrm{m^2/yr}$ of injected aerosol surface area,
\begin{equation}
\eta_S=\int \frac{\rho_pD}{6} F(D)\rho(D)d(\text{log}D).
\end{equation}
The value of $\eta_S$ will directly relate to the total amount of surface area available for heterogeneous reactions, including uptake of species which affect the ozone layer, and is thus a key parameter of interest for SAI safety. Compared to $\eta_M$, the enhanced contribution of smaller particles to the total surface area for a given aerosol burden skews $\eta_S$ to favor larger particles, around $500\,\mathrm{nm}$ for silica. 

Finally, from the perspective of aircraft constraints discussed previously, we can define the air consumption efficiency in $\mathrm{W/m^2}$ per Tg/yr of air,
\begin{equation}
\eta_A=\eta_M/a_R.
\end{equation}

\begin{figure}[!bt]
    \centering
    \includegraphics[width=1\linewidth]{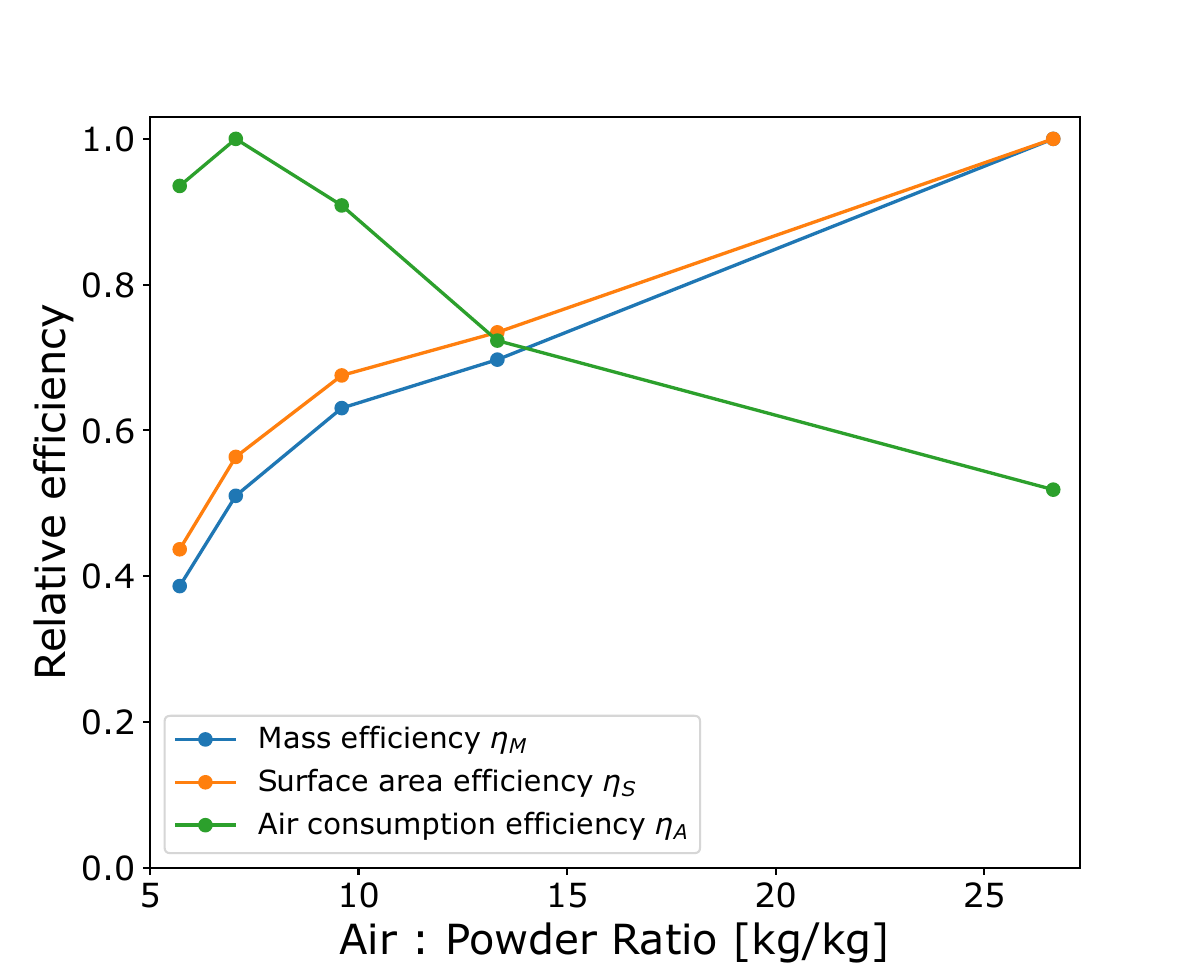}
    \caption{Relative efficiency indices for radiative forcing per annually injected particle mass, surface area, and air, as a function of air to powder mass ratio of a given example dispersal system configuration. Changes in dispersal parameters change the PSD, leading to different, and even opposite trends. All curves have been normalized by their maximum values.  PSD based on performance for 7 bar, room temperature motive air.}
    \label{fig:efficiencies}
\end{figure}

Figure \ref{fig:efficiencies} presents the variation of these three efficiency parameters for a given dispersal system using D500 particles when varying only the air consumption rate; for clarity, all three parameters have been arbitrarily rescaled. As can be seen, the mass efficiency, $\eta_M$, increases with higher air consumption due to improved deagglomeration. The surface area efficiency, $\eta_S$, similarly increases with improved deagglomeration, but at a different rate. For the air consumption efficiency, $\eta_A$, an opposite trend is observed: efficiency decreases with higher air consumption, meaning that the added air flow outweighs the particle mass reduction from improved scattering efficiency. Within the parameter range presented, an approximate reduction of 50\% in material mass and surface area would require an increase of 100\% in air consumption, which would translate accordingly to a larger fleet size if this were the technical constraint. No single presumption can be made about the balancing needed between these trends; however, this observation underscores the need to account for multiple such considerations when analyzing SAI systems, as efficacy, safety, and practicality may strongly compete.

\section{Conclusions}
In this study, we presented how submicron particles can be efficiently dispersed in a size range which is useful for SAI, a challenge which has been recognized as a key risk for the use of solid aerosols for this application. We identify two primary enabling parameters for efficient dispersal. The first requirement is the use of a fabrication method yielding  a narrow monomer size distribution. The specific Stöber sol-gel technique we employed is tunable, suitable for mass-production, and produces almost perfectly spherical particles, thus avoiding the production of monomers which are outside the effective size range. 

The second requisite for efficient dispersal of the particles in this study was the application of a surface treatment to substantially decrease the adhesion energy between the particles. By treating the particle surfaces with silane, replacing the strongly polar hydroxyl terminations with a hydrophobic layer of methyl groups, we drastically improved their dispersibility compared to untreated samples. Combining a theoretical model for particle adhesion forces with surface roughness measurements using AFM, we find a strong correlation between the measured dispersibility and the calculated adhesion, supporting our understanding of the added benefit of surface treatment and roughness. Therefore, although spherical silica particles were used in this study, we note that our findings are not restricted to this material, and may be applicable to any material with similar surface contributions, including composite particles with hydrophobic silica shells encasing other cores\cite{Amyad}. 

For the dispersal system and the particles to be applicable for SAI, the dispersing platform must be able to supply the air needed for deagglomeration without restricting the dispersal to rates which would necessitate an unreasonably large number of aircraft. We show that with the quantities of air that can be supplied from typical turbine engines used in aviation, a system comprising parallel eductors can efficiently disperse particles at a rate which is scalable to millions of tons per year using a fleet on the order of hundreds of planes. We also demonstrate experimentally that air requirements can be substantially reduced by using hot dispersal air, multiple deagglomeration stages, optimally sized monomers, or the addition of a minute fraction of mechanofused nanometric particles. We thus suggest that, for applicable solid particles, pneumatic dispersal can be an effective and efficient building block for SAI technology.

\section{Acknowledgments}
The authors thank Jesse Capecelatro of the University of Michigan and Vicki H.~Grassian of the University of California, San Diego for useful discussions. E.H.~also thanks David Hunter and Mark Young for helpful discussions. 

\section{Disclosure}
Y.S., E.Y.L., Y.B.Y., O.A., E.L., S.R., A.L.~and A.S.~are affiliated with Stardust Labs Ltd., a sunlight reflection technology development company.

\begingroup
\setlength{\emergencystretch}{1.05em} 
\printbibliography
\endgroup

\newpage
\appendix
\section{Supplementary Materials}
\subsection{Surface characterizations}
\label{surface characterization}

For AFM (Atomic Force Microscopy) analyses, Au-coated Si wafers (Platypus Technologies) were prepared by cleaning in warm H$_2$O$_2$  ($\sim60\,^\circ$C) and room temperature isopropanol followed by a rinse with milliQ water. AFM probes used in this study along with the pre-cleaned wafers were further cleaned using the UV/O$_3$ Pro-cleaner (Bioforce Nanosciences) for 30 min to remove any residual organics. Individual particles were fixed onto substrates using a thin layer of epoxy resin. The epoxy resin was allowed to dry prior to AFM measurements.

Particle samples were analyzed using a commercial nanoIR2 microscopy system (Anasys Instruments, Santa Barbara, CA, USA). The instrument consists of an atomic force microscope integrated with a pulsed, tunable infrared optical parametric oscillator (OPO) laser source with repetition rate of 1 kHz, tuning range from 850 to 1800 cm$^{-1}$ and 2500 to 3800 cm$^{-1}$ and an average spectral resolution of 4 cm$^{-1}$. All analyses were conducted in an environmental chamber maintained below 3\% relative humidity and a temperature of 298 K. AFM-PTIR (Atomic Force Microscopy – Photothermal Infrared) spectra were used to confirm lack of epoxy residues on the substrate-fixed particles. 

Chemical maps of particles were collected at 3700 cm$^{-1}$ for surface hydroxyl groups using a HQ:CSC-17 Cr/Au high sensitivity AFM probe from MikroMasch, at a scan rate of 0.1 Hz, polarization of $90^\circ$, data point spacing resolution of 1.559 cm$^{-1}$/pt, averaging 512 laser pulses per wavenumber, and 2x co-averages were used. 

RMS surface roughness and information about surface asperity were extracted from AFM height images collected in contact mode with the Gwyddion 2.56 software. These images were collected at 0.4 Hz scan rate. 

Measurements of force of adhesion were conducted using an antimony doped Si probe with a native SiO$_2$ layer. The probe (Bruker corporation) had a 20 nm tip radius, 0.2 N/m spring constant and 13 kHz frequency using the force controls module. Approach and retreat curves were collected with 0.05 Hz rate at 1000 pt resolution, 500 nm Z-range and in relative limit mode with deflection delta of 0.4 V. The Si probe was calibrated using a clean Si wafer substrate with a native oxide layer. AFM height images in contact mode were collected before and after force curves measurements. Force of adhesion was extracted from the retreat curves\cite{yuan2025measurement}. Between 15 – 20 measurements per material across individual particles were collected and averaged along with a reported standard deviations.

\newpage
\subsection{Experimental Setup}

\begin{figure}[!b]
    \centering
    \includegraphics[width=0.5\linewidth]{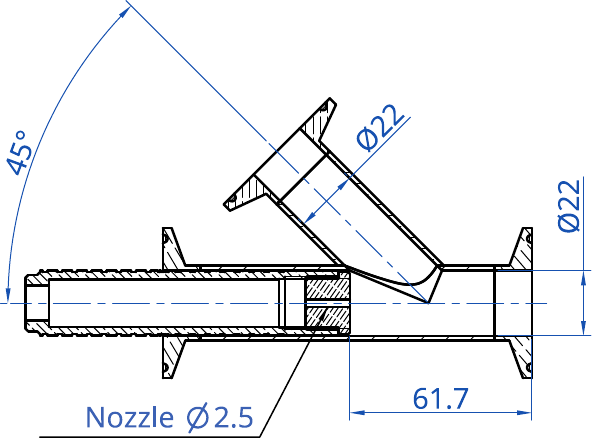}
    \caption{Eductor used in the dispersal setup. Pressurized motive air enters from the left, after passing through an optional air heater. Suction, created by supersonic expansion through the nozzle, pulls particle-laden air in from the top. Turbulent shear develops after the mixing between the flows, and the particles are dispersed as a plume or fed into additional stages to the right. Dimensions are in millimeters.}
    \label{fig:eductor}
\end{figure}

\begin{figure}[!b]
    \centering
    \includegraphics[width=0.75\linewidth]{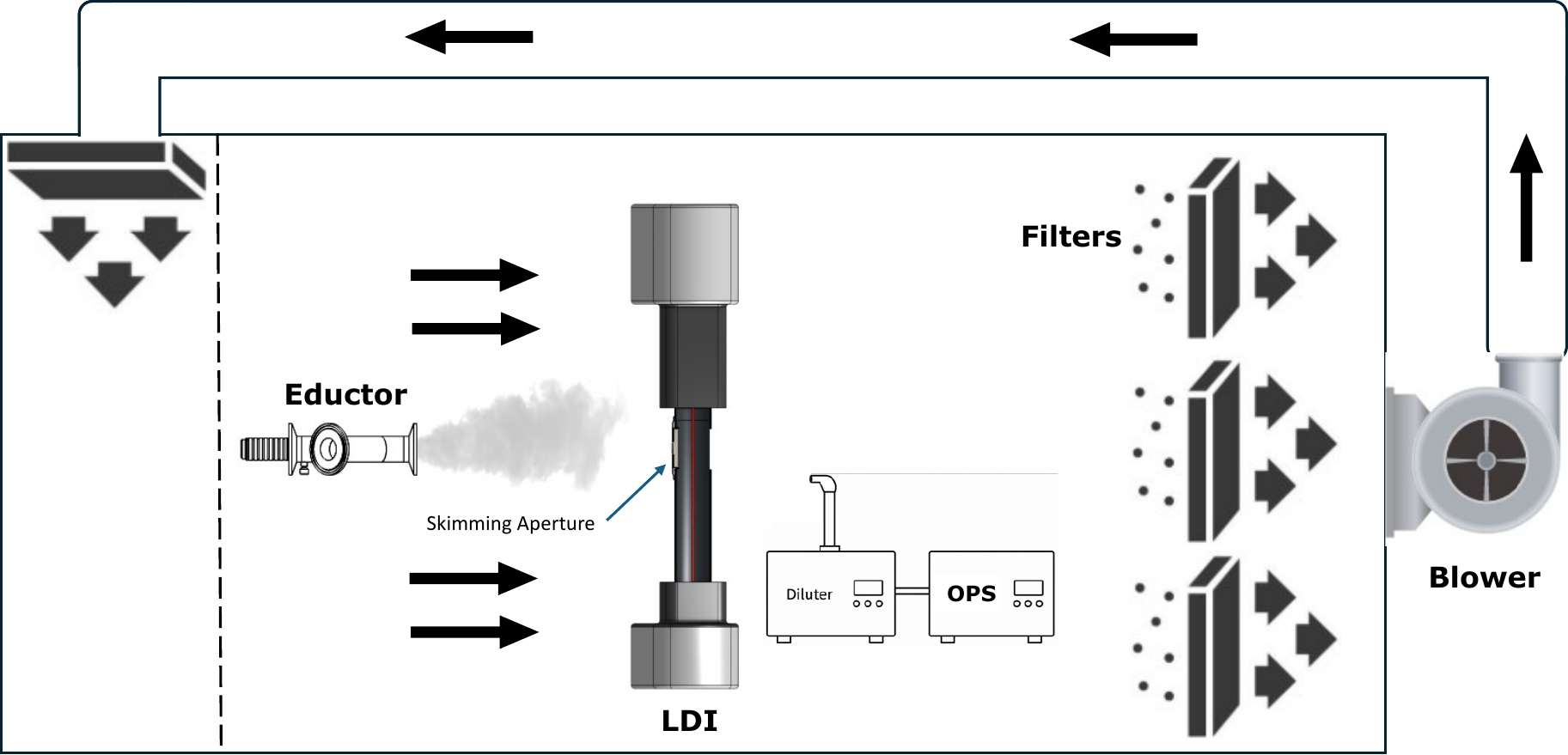}
    \caption{Spray tunnel facility. The dispersal system is placed up to 4 meters away from the particle sampling bench, which includes a laser diffraction imaging (LDI) instrument and an optical particle spectrometer (OPS) sampling through a calibrated dilution system (diluter). The air is circulated and filtered to avoid build-up of particles near the instruments. The experiment is monitored and operated remotely from an adjacent control room.}
    \label{fig:spray tunnel}
\end{figure}

\end{document}